\documentclass[prd,superscriptaddress,showpacs,nofootinbib,amsmath,amssymb,aps,10pt]{revtex4}

\usepackage{bm}
\usepackage{amsfonts}
\usepackage{latexsym}
\usepackage[latin1]{inputenc}
\usepackage{graphicx}
\usepackage{amsmath}
\usepackage{palatino}
\usepackage{mathpazo}
\usepackage{booktabs}
\usepackage{dcolumn}

\usepackage{wrapfig}
\usepackage{wasysym}

\def\jnl@style{\it}
\def\aaref@jnl#1{{\jnl@style#1}}

\def\aaref@jnl#1{{\jnl@style#1}}

\def\aj{\aaref@jnl{AJ}}                   
\def\apj{\aaref@jnl{ApJ}}                 
\def\apjl{\aaref@jnl{ApJ}}                
\def\apjs{\aaref@jnl{ApJS}}               
\def\apss{\aaref@jnl{Ap\&SS}}             
\def\aap{\aaref@jnl{A\&A}}                
\def\aapr{\aaref@jnl{A\&A~Rev.}}          
\def\aaps{\aaref@jnl{A\&AS}}              
\def\mnras{\aaref@jnl{Mon.~Not.~Roy.~Astron.~Soc.}}             
\def\prd{\aaref@jnl{Phys.~Rev.~D}}        
\def\prc{\aaref@jnl{Phys.~Rev.~C}}  
\def\prl{\aaref@jnl{Phys.~Rev.~Lett.}}    
\def\qjras{\aaref@jnl{QJRAS}}             
\def\skytel{\aaref@jnl{S\&T}}             
\def\ssr{\aaref@jnl{Space~Sci.~Rev.}}     
\def\zap{\aaref@jnl{ZAp}}                 
\def\nat{\aaref@jnl{Nature}}              
\def\aplett{\aaref@jnl{Astrophys.~Lett.}} 
\def\apspr{\aaref@jnl{Astrophys.~Space~Phys.~Res.}} 
\def\physrep{\aaref@jnl{Phys.~Rep.}}      
\def\physscr{\aaref@jnl{Phys.~Scr}}       
\def\commat{\aaref@jnl{Comm.~Math.~Phys.}}              
\def\science{\aaref@jnl{Science}}               
\def\cqg{\aaref@jnl{Classical Quant.~Grav.}}            
\def\jpcs{\aaref@jnl{JPCS}}                                     
\def\ijmpd{\aaref@jnl{Int.~J.~Mod.~Phys.~D}}                    
\def\grg{\aaref@jnl{Gen.~Relat.~Gravit.}}               
\def\rpp{\aaref@jnl{Rep.~Prog.~Phys.}}          
\def\npa{\aaref@jnl{Nucl.~Phys.~A}}        
\def\lrr{\aaref@jnl{Living Rev.~Rel.}}                   
\def\jcap{\aaref@jnl{J.~Cosmology Astropart.~Phys.}}    
\def\rmp{\aaref@jnl{Rev.~Mod.~Phys.}}   

\allowdisplaybreaks[1]

\begin{document}

\title{The I-Q relations for rapidly  rotating neutron stars in $f(R)$ gravity}

\author{Daniela D. Doneva}
\email{daniela.doneva@uni-tuebingen.de}
\affiliation{Theoretical Astrophysics, Eberhard Karls University of T\"ubingen, T\"ubingen 72076, Germany}
\affiliation{INRNE - Bulgarian Academy of Sciences, 1784  Sofia, Bulgaria}

\author{Stoytcho S. Yazadjiev}
\email{yazad@phys.uni-sofia.bg}
\affiliation{Department of Theoretical Physics, Faculty of Physics, Sofia University, Sofia 1164, Bulgaria}
\affiliation{Theoretical Astrophysics, Eberhard Karls University of T\"ubingen, T\"ubingen 72076, Germany}

\author{Kostas D. Kokkotas}
\email{kostas.kokkotas@uni-tuebingen.de}
\affiliation{Theoretical Astrophysics, Eberhard Karls University of T\"ubingen, T\"ubingen 72076, Germany}


\begin{abstract}
In the present paper we study the behavior of the normalized $I$-$Q$ relation for neutron stars in a particular class of $f(R)$ theories of gravity, namely the $R^2$ gravity that is one of the most natural and simplest extensions of general relativity in the strong field regime. We study both the slowly and rapidly rotating cases. The results show that the $I$-$Q$ relation remain nearly equation of state independent for fixed values of the normalized rotational parameter, but the deviations from universality can be a little bit larger compared to the general relativistic case. What is the most interesting in our studies, is that the differences with the pure Einstein's theory can be large reaching above 20\%. This is qualitative different from the majority of  alternative theories of gravity, where the normalized $I$-$Q$ relations are almost indistinguishable from the general relativistic case, and can lead to observational constraints on the $f(R)$ theories in the future.
\end{abstract}

\maketitle
\date{}
\section{Introduction}
Studies of generalized theories of gravity are becoming more and more intense in the last decade. There are both theoretical and observational motivations for this. On one hand the theories trying to unify all the interactions predict that the standard Einstein-Hilbert action should be modified. On the other hand it was shown in many cases that studying generalizations of Einstein's gravity can give us a deeper understanding of general relativity (GR) itself. On the observational front still remain phenomena that do not fit very well in the standard framework, such as the accelerated expansion of the universe, and that is why modifications of the theory of gravity are often employed as an alternative explanations. One should keep also in mind, that even though general relativity is very well tested in the weak field regime, the strong field remains essentially unconstrained that leaves space for a variety of modifications.

One of the most natural generalizations of Einstein's theory of gravity are the $f(R)$ theories, where the Ricci scalar $R$ in the Einstein-Hilbert action is replaced by some function of $R$. Such modification has a theoretical motivation for example from the quantum field theory in curved spacetime.
$f(R)$ theories are also widely used as an alternative explanation of the dark energy phenomena which places them amongst the most popular and widely explored alternative theories of gravity. Most of the studies on $f(R)$ theories though are in cosmological aspect in relation to the accelerated expansion of the universe \cite{Sotiriou2010,Nojiri2011,Capozziello2011}. The examinations of the astrophysical manifestations of these theories is more scarce and this would be the main focus of our paper.

Natural objects to study within the $f(R)$ theories of gravity at astrophysical scales are the neutron stars, where the strong gravity effects are non-negligible.  The neutron stars
within the $f(R)$ theories  can differ significantly from their GR counterpart \cite{Yazadjiev2014,Yazadjiev2015b,Staykov2014} which makes them a very good candidate to test $f(R)$ theories on astrophysical scales. Unfortunately one has to pay a high price -- the nuclear matter equation of state (EOS) at densities as high as the ones in the neutron star cores, is still unknown. That is why in many cases the deviations coming from the generalizations of Einstein's gravity are comparable or even smaller than the deviations resulting from the uncertainties in the EOS.

A way to circumvent this problem is to search for predictions or derive relations that are independent of the EOS. This was exactly the idea in \cite{Yagi2013,Yagi2013a} where the famous $I$-Love-$Q$ relations were discovered. These relations connect the normalized neutron star moment of inertia $I$, quadrupole moment $Q$ and the tidal Love number $\lambda$. They were later explored in the regime of rapid rotation \cite{Doneva2014,Pappas2014,Chakrabarti2014}, large tidal deformations \cite{Maselli2013}, in the presence of magnetic fields \cite{Haskell2013}, for higher multipole moments \cite{Yagi2014} and in the presence of anisotropic pressure \cite{Yagi2015}. We should note that also a variety of other (nearly) EOS independent relations exists in the literature connecting different neutron star properties, their oscillations spectrum, etc. (see for example \cite{Lattimer2001,Urbanec2013,Baubock2013,AlGendy2014,Andersson98a,Tsui2005,Gaertig10,Doneva2013a,Delsate2015,Pani2015,Pappas2015b}).

The most important property of the $I$-Love-$Q$ relations is that they are practically independent of the EOS (for moderate magnetic fields). Several applications were proposed and one of the most important is braking the degeneracy between the spins and the quadrupole moment of neutron star inspirals, and testing alternative theories of gravity. In connection to the latter, the $I$-Love-$Q$ relations were examined in a variety of alternative theories of gravity, such as the dynamical Chern-Simons gravity \cite{Yagi2013,Yagi2013a}, Eddington-inspired Born-Infeld gravity \cite{Sham2014}, Einstein-Gauss-Bonnet-dilaton theory \cite{Kleihaus2014} and scalar-tensor theories of gravity \cite{Pani2014,Doneva2014a,Pappas2015,Pappas2015a}. In all of these theories the $I$-Love-$Q$  relations are pretty much EOS independent. With the exception of the dynamical Chern-Simons gravity, the resulting relations are quite similar to the GR case and they can not be used to test the alternative theories of gravity\footnote{One should keep in mind that this is true only for the normalized relations. The unnormalized quantities can deviate significantly and potentially lead to some observational effects.}.

The purpose of the present paper is to explore the normalized $I$-$Q$ relations for $f(R)$ theories of gravity and more specifically for the $R^2$ gravity, where $f(R) = R+aR^2$. We will examine both the slowly and rapidly rotating regime. The main goals is to determine if the relations are still EOS independent and whether significant deviations from GR can be observed. We will use the fact that $f(R)$ theories are mathematically equivalent to a particular class of scalar-tensor theories with a nonzero potential of the scalar field. As we will demonstrate below, exactly the presence of such nonzero potential leads to interesting results and makes the problem (and the conclusions) qualitative different from the one considered in \cite{Doneva2014a}.

\section{Basic equations}
In our calculations we will use the fact that $f(R)$ theories are mathematically equivalent to a specific class of scalar-tensor theories. Here we will give very briefly the main points and we refer the reader to \cite{Yazadjiev2014,Yazadjiev2015b,Staykov2014} where the problem is discussed in detail.

The $f(R)$ gravity action can be written in the following general form
\begin{eqnarray}\label{A}
	S= \frac{1}{16\pi G} \int d^4x \sqrt{-g} f(R) + S_{\rm
		matter}(g_{\mu\nu}, \chi),
\end{eqnarray}
where $R$ is the Ricci scalar curvature with respect to the space-time metric $g_{\mu\nu}$, $S_{\rm matter}$ is the action of the matter, and the matter fields are denoted by $\chi$. This action is mathematically equivalent to the following scalar-tensor theory (STT) action
\begin{eqnarray}\label{EFA}
	S=\frac{1}{16\pi G} \int d^4x \sqrt{-g^{*}}\left[ R^{*} - 2
	g^{*\mu\nu}\partial_{\mu}\varphi \partial_{\nu}\varphi - V(\varphi)
	\right] + S_{\rm
		matter}(A^2(\varphi)g^{*}_{\mu\nu},\chi),
\end{eqnarray}
where the coupling function $A(\varphi)$ and the scalar-field potential $V(\varphi)$ have the following form
\begin{eqnarray}\label{AV}
A^2(\varphi)=e^{-\frac{2}{\sqrt{3}}\varphi}, \,\,\, V(\varphi)=A^4(\varphi)\left(R \frac{df}{dR} - f(R)\right).
\end{eqnarray}
Here $R^{*}$ is the Ricci scalar curvature with respect to the metric $g^{*}_{\mu\nu}$ and  $\varphi$ is the scalar field. In this paper we will concentrate on the so-called $R^2$ gravity defined by $f(R) = R + aR^2$. In this case the scalar-field potential takes the form
\begin{eqnarray}\label{AV_R2}
V(\varphi)= \frac{1}{4a}
\left(1-e^{-\frac{2\varphi}{\sqrt{3}}}\right)^2.
\end{eqnarray}

The STT action \eqref{EFA} is written in the so-called Einstein frame, that is not the physical one and it is introduced because it simplifies the field equations substantially. The physical quantities, such as mass, distance, etc., are measured in the Jordan frame, where the two frame are connected by a conformal transformation of the metric and a redefinition of the scalar field. We will not go into details and we refer the reader to  \cite{Doneva2013,Yazadjiev2014} for an extensive discussion of the problem.

We will consider stationary, axisymmetric and asymptotically flat solutions of the field equations describing rotating compact starts. Thus the Einstein frame spacetime metric can be presented in the following form
\begin{eqnarray}
ds_{*}^2 = - e^{2\nu}dt^2 + \rho^2 B^2 e^{-2\nu}(d\phi - \omega dt)^2 + e^{2\zeta - 2\nu}(d\rho^2 + dz^2),
\end{eqnarray}
where all the metric functions depend on the coordinates $\rho$ and $z$ only. Let us comment on the asymptotic behavior of the metric functions and the scalar field. It is derived from the reduced field equations that can be found in \cite{Yazadjiev2015b}. It is more convenient to present the asymptotic behaviour in the quasi-isotropic coordinates $r$ and $\theta$ defined by

\begin{eqnarray}
\rho=r\sin\theta , \; \; z=r\cos\theta.
\end{eqnarray}

Using these coordinates and keeping only terms up to the order of $r^{-3}$, one can obtain
\begin{eqnarray}
&&\nu \approx - \frac{M}{r} + \left[ \frac{b}{3} + \frac{\nu_2}{M^3} P_{2}(\cos\theta)\right] \left( \frac{M}{r}\right)^3 , \label{ASMPT1}\\
&&B\approx 1 + b \left( \frac{M}{r}\right)^2, \label{ASMPT2}\\
&&\omega \approx \frac{2J}{r^3} , \label{ASMPT3}\\
&& \zeta\approx -\left\{\frac{1}{4}
+ \frac{1}{3}\left[b + \frac{1}{4}\right]\left[1 - 4P_{2}(\cos\theta)\right] \right\} \left( \frac{M}{r}\right)^2 , \label{ASMPT4}
\end{eqnarray}
where $M$ and $J$ are the mass and the angular momentum, $b$, $\nu_2$ and $\varphi_{2}$ are constants and $P_{2}(\cos\theta)$ is the second Legendre polynomial. The scalar field on the other hand decreases exponentially at infinity and one can show that
\begin{eqnarray}\label{eq:phi_asympt}
	\varphi \sim \frac{e^{-m_\varphi r}}{r},
\end{eqnarray}
for large values of $r$, where $m_\varphi$ is the scalar field mass equal to $m_\varphi=1/(\sqrt{6}a)$. 
One can make two important conclusions from eq. \eqref{eq:phi_asympt}. First, the scalar charge of the scalar field is zero\footnote{The scalar charge in usually defined as the coefficient in front of the $1/r$ term in the scalar field expansion at infinity.} and therefore it does not contribute to the asymptotic of the metric functions \eqref{ASMPT1} -- \eqref{ASMPT4} contrary to the case of the scalar-tensor theories (with massless scalar field) admitting scalarization \cite{Doneva2014a}. Second, the exponential decay of the scalar field is controlled by the parameter $a$ and smaller $a$ corresponds to more rapid decay. Thus the general relativistic case corresponds to the limit $a\rightarrow 0$.

Since the scalar field does not contribute to the asymptotic of the metric, the quadrupole moment would have the same form as in general relativity. Thus one can  obtain \cite{Stergioulas2011,Pappas2012}
\begin{eqnarray}\label{eq:QuadrupoleMoment}
Q=- \nu_{2} - \frac{4}{3}\left[b + \frac{1}{4}\right]M^3.
\end{eqnarray}

In our calculations we will use also the moment of inertia $I$ defined in the usual way
\begin{eqnarray}\label{eq:MomentOfInertia}
I=\frac{J}{\Omega}
\end{eqnarray}
with $\Omega$ being the angular velocity of the star.

Equations \eqref{eq:QuadrupoleMoment} and \eqref{eq:MomentOfInertia} are written in the Einstein frame. Therefore, one has to transform them in the physical Jordan frame. But,
taking into account the exponential decay of scalar field,  one can easily show that the moment of inertia and the quadrupole moment are the same in both frames for the particular scalar-tensor theory and therefore for the $R^2$ gravity we consider in the present paper. Therefore, formulae \eqref{eq:QuadrupoleMoment} and \eqref{eq:MomentOfInertia} are valid also in the physical Jordan frame.

In our calculations we use the dimensionless parameter $a\to a/R^2_{0}$, where $R_{0}$ is one half of the solar gravitational radius   $R_{0}=1.47664 \,{\rm km}$ (i.e. the solar mass in geometrical units).

\section{Numerical results}
In our calculations we employed four equations of state (EOS) that cover a wide range of stiffness --  APR, SLy4, FPS and Shen2D (the zero temperature limit of the Shen EOS). The first two EOS are considered as standard since they are above the two solar mass barrier and have radii in the preferred range according to the observations \cite{Lattimer12,Steiner2010,Demorest10,Antoniadis13}. The FPS EOS is softer with maximum mass below the $2M_\odot$. The Shen2D EOS on the other hand is stiff and has high maximum mass, but the typical radii are around 14-15km that is larger compared to the observational estimates. Even though EOS FPS and Shen2D do not fit all the observations well, we have chosen them in order to cover a larger range of stiffness and to check up to what extend our relations are EOS independent. The rotating neutron star solutions in $R^2$ gravity are obtained using an extended version of the rns code \cite{Yazadjiev2015b}.

The goal of the present paper is to study the relation between the moment of inertia $I$ and the quadrupole moment $Q$ of neutron stars in $f(R)$ theory in connection to the famous $I$-Love-$Q$ relations. In order to obtain (nearly) equation of state independent relations, $I$ and $Q$ have to be properly normalized. A natural choice is ${\bar I} \equiv I/M^3$ and ${\bar Q} \equiv Q/(M^3 \chi^2)$, where $\chi\equiv J/M^2$. Since we are considering rotating models, we will have two main goals -- first is to determine what the deviations from pure general relativity are, and second -- to check up to what extend the relations are equation of state independent. We should note that a different normalization might make the relations more or less universal and increase or decrease the differences between the gravitational theories. A conjecture was made in \cite{Yazadjiev2015a} that always exists a normalization that can make a given relation universal. But one should not choose aways the most universal normalization, but instead the most relevant one from both physical and observational point of view. The particular normalization in the current paper is the standard one used in all the previous papers on $I$-Love-$Q$ relations, including relations in alternative theories of gravity. It has a number of physical justifications and makes the comparison of the results much easier.

\begin{figure}[]
	\centering
	\includegraphics[width=0.48\textwidth]{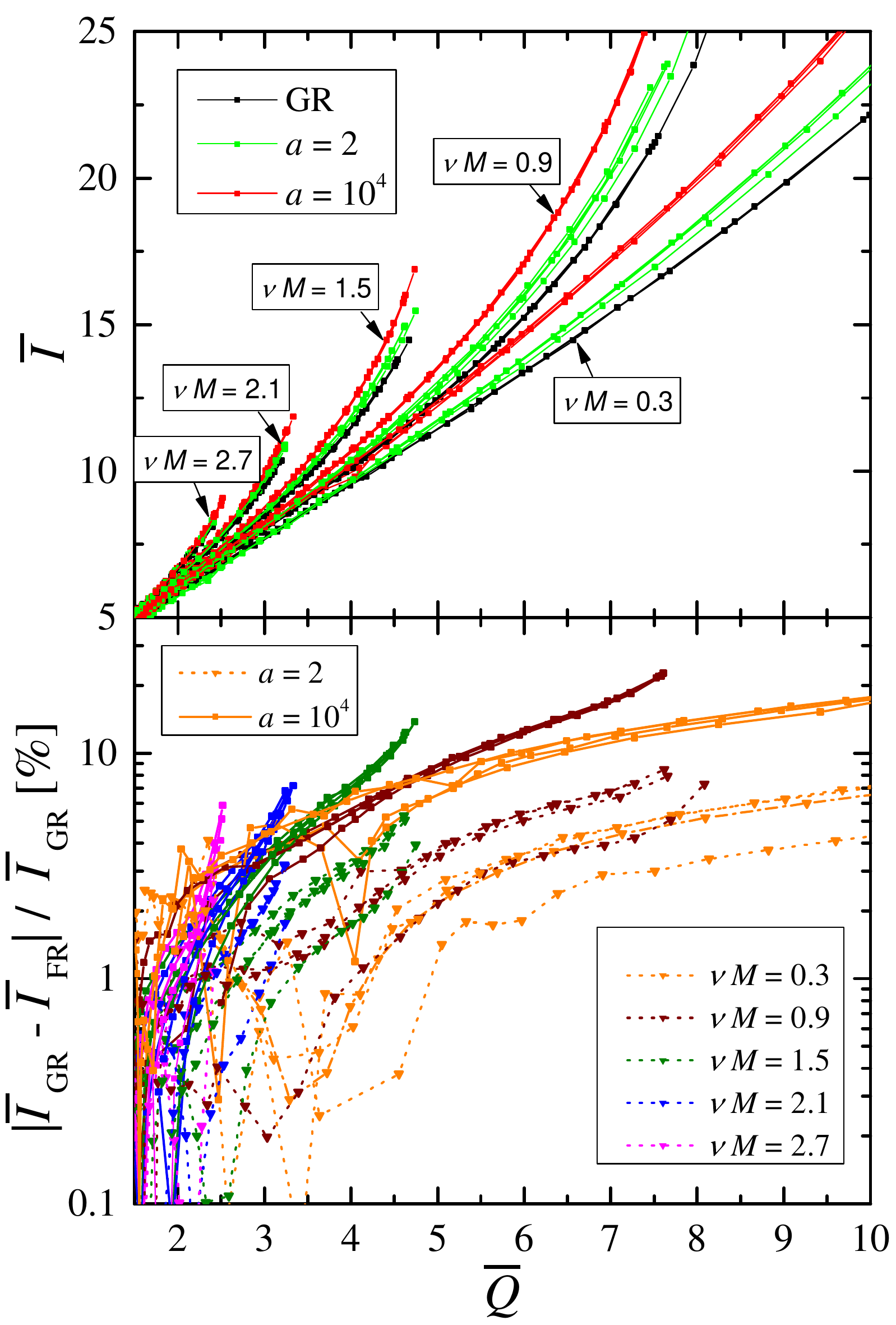}
	\includegraphics[width=0.48\textwidth]{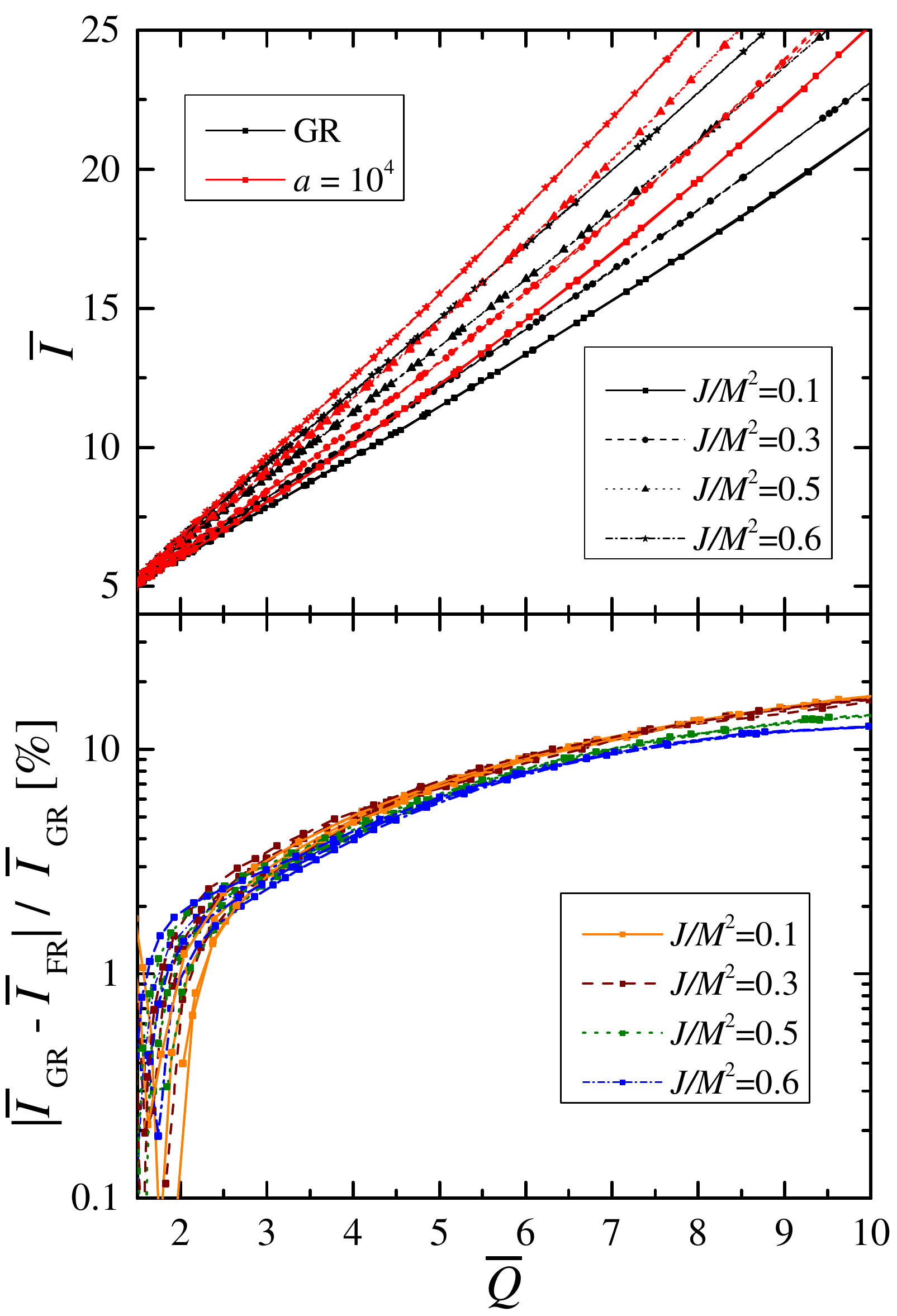}
	\caption{$\bar{I}-\bar{Q}$ relation for sequences with fixed rotational parameter $\nu M$ (right panel) and fixed $J/M^2$ (left panel). The lower panel for both figures represents the relative deviation of the results in $f(R)$ gravity from the GR case.} \label{Fig:IQ}
\end{figure}

We work with two different rotational parameters that are popular in the literature -- the normalized rotational frequency $\nu M$ ($\nu$ is the rotational frequency of a star in kHz, $M$ is the mass in solar masses) and the normalized angular momentum $J/M^2$ ($J/M^2$ is dimensionless in units $c=G=1$). The ${\bar I}-{\bar Q}$ relations for sequences of neutron star models with constant value of $\nu M$ are presented in the left panel of Fig. \ref{Fig:IQ} and with constant $J/M^2$ -- in the right panel of Fig. \ref{Fig:IQ}. The sequences range from the slow rotation limit ($\nu M=0.3$ and $J/M^2=0.1$) to very fast rotation ($\nu M=2.7$ and $J/M^2=0.6$). For every single $\nu M$ three cases are shown -- the general relativistic limit ($a=0$) and $R^2$ gravity with $a=2$ and $a=10^4$ ($a=10^4$ gives nearly the maximum possible deviation from GR). In the case of $J/M^2$ we show the results only for $a=0$ and $a=10^4$ in order to have a better visibility.

In the bottom part of the plots in Fig. \ref{Fig:IQ} the deviations $\Delta I$ of the $f(R)$ gravity results from the GR case are shown, defined as
\begin{eqnarray}
\Delta I = \frac{\left|{\bar I}_{GR} - {\bar I}_{FR}\right|}{{\bar I}_{GR}}.
\end{eqnarray}
Here ${\bar I}_{GR}$ and ${\bar I}_{FR}$ are the corresponding values of the moment of inertial in the GR and $f(R)$ cases respectively and the plotted values of $\Delta I$ are in percentages.

The most important result is that $\Delta I$ reaches above 20\% for the less massive models (with larger $\bar Q$), slower rotation and large values of $a$. Such deviation is much higher that the one observed for other alternative theories of gravity such as scalar-tensor theories which admit scalarization \cite{Doneva2014a}. This is surprising at first sight because, as we described above, we are using the mathematical equivalence between $f(R)$ theories and scalar-tensor theories in our calculations. But there is a very important difference between the particular class of STT considered here that is equivalent to the $f(R)$ theories (see eqs. \eqref{AV}), and the STT considered in \cite{Doneva2014a,Doneva2013}. It steams from the fact that the scalar-field mass  in our case is nonzero which leads to its exponential decay (see eq. \eqref{eq:phi_asympt}). Thus the scalar field has a zero scalar charge and it does not contribute directly to the asymptotic of the metric functions at infinity and the quadrupole moment. Therefore eqs. \eqref{ASMPT1} and \eqref{eq:QuadrupoleMoment} are valid both in GR and $R^2$ gravity.

This is not the case with the STT considered in \cite{Doneva2014a,Doneva2013}, which admits spontaneous scalarization and has a massless scalar field. A nonzero scalar charge is present there which enters explicitly in the equation for the quadrupole moment and roughly speaking it compensates the deviations from GR in the  ${\bar I} - {\bar Q}$ relation\footnote{We should note that only the normalized ${\bar I} - {\bar Q}$ relation is close to GR for the STT considered in \cite{Doneva2014a}. The unnormalized quantities $I$ and $Q$ can reach very large deviations from GR especially in the rapidly rotating case}. Such ``compensation'' of the scalar field effect does not exist in the STT considered in this paper (and therefore in $f(R)$ gravity) and that is why large differences from GR can be observed.

The deviations from EOS universality are small, below roughly 1\%, for the general relativistic case (a=0) and for the case of very large $a$. The deviations can increase for small $a$ reaching roughly 3\%, but this case is not so interesting  since it gives small differences with GR. Another point is that deviations as large as 3\% are observed only for more extreme EOS, such as the Shen2D EOS. For the ``standard'' modern realistic EOS the deviations from EOS universality are still within roughly $1\%$.

Let us comment on the effect of rotation. From the bottom panel in Fig. \ref{Fig:IQ} it is evident that for fixed values of ${\bar Q}$ the deviations from GR are comparable for different values of the normalized rotational parameters $\nu M$ and $J/M^2$. Therefore, large differences can be observed already for slow rotation that is very important since a big portion of the observed systems have rotational frequencies up to a few hundred Hz. If the sequences with fixed normalized rotational parameter terminate at smaller values of ${\bar Q}$, i.e. at large masses, the deviations from GR are relatively small that naturally come from the fact that they are close to the black hole limit  where the ${\bar I} - {\bar Q}$ relation does not depend on the internal structure of the star (this is observed in Fig. \ref{Fig:IQ} especially in the case of fixed $\nu M$). The deviations from EOS universality are not sensitive to the particular value of $\nu M$ or $J/M^2$, i.e. they are of the same order for all of the sequences.

Another important point one can notice is that the effect of rotation is qualitatively the same as the effect of increasing the parameter $a$. Therefore, in order to test the deviations from general relativity we should either have additional information about the rotational rate of the star from independent observations, or deal with objects that are in general slowly rotating (the ${\bar I} - {\bar Q}$ dependences are almost indistinguishable for frequencies less than a few hundred Hz). The opposite line of conclusions is also possible -- it would be very difficult to test any deviations from slow rotation via the $I$-Love-$Q$ relations since certain modifications of gravity lead to similar results.

\section{Conclusions}
In the present paper we have studied the normalized ${\bar I} - {\bar Q}$ dependence for rotating neutron stars in $R^2$ theory of gravity. Both the slowly and the rapidly rotating cases are examined. In our calculation we made use of the fact that the $f(R)$ theories are mathematically equivalent to a particular class of scalar-tensor theories with a nonzero potential of the scalar field. The effective mass of the scalar field in this scalar-tensor representation is backproportional to the $R^2$ gravity parameter $a$ and the scalar field decays exponentially at infinity. Therefore the scalar charge is zero and the asymptotic of the metric at infinity and the quadrupole moment have the same form as in pure general relativity (contrary to other classes of STT with massless scalar field \cite{Doneva2014a}). The rotating neutron star solutions were obtained with an extended version of the rns code \cite{Yazadjiev2015b}.

The most important conclusion from our results is that the dependences can deviate significantly from the GR case and the differences can reach above 20\% that can be observationally relevant. The deviations are in general larger for larger values of the $R^2$ gravity parameter $a$, smaller neutron star masses and slower rotation. This is qualitatively different from the case of scalar-tensor theories with vanishing potential, where the normalized ${\bar I} - {\bar Q}$ relation differs only marginally from GR even in the rapidly rotating case. A rough explanation of this phenomena comes from the fact that the scalar charge is zero in the scalar-tensor representation of the $R^2$ gravity and thus it has no explicit contribution to the quadrupole formula. The effect of $R^2$ gravity is qualitatively similar to the effect of changing the normalized rotational parameter $\nu M$ or $J/M^2$. So one should be careful when trying to distinguish between the two effects in the future observations.

The  ${\bar I} - {\bar Q}$ relations in $R^2$ gravity remain nearly EOS independent for fixed values of the normalized rotational parameters $\nu M$ and $J/M^2$ similar to the pure Einstein's theory. The deviations are typically of the order of $1\%$ for large values of $a$ or large masses and they increase up to roughly $3\%$ with the decrease of $a$ and $M$. This is certainly below the expected observational accuracy. Also deviations as large as 3\% are observed only for more extreme EOS, and therefore they are of limited interest.

It is interesting to note that the considered $f(R)$ theories are one of the few generalized theories of gravity that lead to large deviations from GR as far as the ${\bar I} - {\bar Q}$ relations are concerned. As we commented above, this is due to the fact that the scalar field in the scalar-tensor representation of the $R^2$ gravity is massive (i.e. has a finite range). Hence we expect that other classes of massive scalar-tensor theories will have similar behavior and such a study is underway.

At the end we will briefly comment on the observational perspectives. As we have already pointed out, larger deviations in the ${\bar I} - {\bar Q}$ relations are observed for lower masses and slow rotation. But this is exactly the case of merging neutron stars that are supposed to have typical masses around $1.4 M_\odot$ and to be slowly rotating due to the long evolution time. Probably the $f(R)$ gravity will also change the gravitational wave signal emitted by such inspiraling neutron stars similar to the case of scalar-tensor theories \cite{Barausse2013,Palenzuela2014,Shibata2014,Taniguchi2015} but additional studies in this direction are needed. One of the most promising application of the $I$-Love-$Q$ relations is in the parameter extraction of such merging binaries and therefore one should keep in mind the possible effects that might come from the alternative theories of gravity and the $f(R)$ theories in particular.

As far as isolated and binary NS are concerned, the precise observation of two of the quantities in the $I$-Love-$Q$ relation would serve as a direct test of GR and can help us to impose constraints on $f(R)$ theories. But we should have at least some information about the rotational frequency of these stars and whether they are slowly or rapidly rotating, since the deviations coming from $R^2$ gravity are qualitative the same as the effect of rotation.

\section*{Acknowledgements}

DD would like to thank the European Social Fund and the Ministry Of Science, Research and the Arts Baden-Württemberg for the support. SY and KS would like to thank the Research Group Linkage Programme of the Alexander von Humboldt Foundation for the support. The support by the Bulgarian NSF Grant DFNI T02/6, Sofia University Research Fund under Grant 70/2015 and "New-CompStar" COST Action MP1304 is gratefully acknowledged.


\bibliography{references}

\end{document}